\documentstyle[epsfig,amsmath,amssymb,graphicx]{elsart}

\def\slash#1{\setbox0=\hbox{$#1$}               
   \dimen0=\wd0                                 
   \setbox1=\hbox{/} \dimen1=\wd1               
   \ifdim\dimen0>\dimen1                        
      \rlap{\hbox to \dimen0{\hfil/\hfil}}      
      #1                                        
   \else                                        
      \rlap{\hbox to \dimen1{\hfil$#1$\hfil}}   
      /                                         
   \fi}                                         %

\def\be{\begin{eqnarray}}
\def\ee{\end{eqnarray}}

\newcommand{\bea}{\begin{eqnarray}}
\newcommand{\eea}{\end{eqnarray}}       

\begin{document}

\begin{frontmatter}

\title{Linking the Quark Meson Model with QCD at High Temperature}


\author{Jens Braun$^{a}$, Kai Schwenzer$^{b}$, Hans--J\"urgen Pirner$^{c}$}
\address{Institut f\"ur Theoretische Physik, Universit\"at Heidelberg, \\
Philosophenweg 19, 69120 Heidelberg, Germany}
\address{(a) jbraun@tphys.uni-heidelberg.de, (b)
  kai@tphys.uni-heidelberg.de, \\
   (c) pir@tphys.uni-heidelberg.de}

\begin{abstract}
We model the transition of  a system of quarks and gluons at high energies
to a system of  quarks and mesons at low energies 
in a consistent renormalization group approach. Flow equations
interpolate between the physics of the high-temperature degrees of freedom and the low-temperature dynamics at a scale of $1$ GeV. We
also discuss the dependence of the equation of state on baryon density
and compare our results with recent lattice gauge simulations.
\end{abstract}

\end{frontmatter}

The equation of state of QCD in extreme conditions of high temperature
and/or high baryon density is at the focus of  
interest in heavy-ion collisions and astrophysics, i.e.  
neutron stars and supernovae.
The current heavy-ion experiments at SPS (CERN) and RHIC (BNL) have
challenged theoretical work on strong interactions at finite
temperatures, due to the possible observation of new phases of
matter. In particular, one expects a phase transition which separates a
low-temperature region of hadrons from a high-temperature region of
quarks and gluons, where both confinement is absent and chiral symmetry
is restored. 
During the last years various direct numerical lattice simulations of QCD at
finite temperature and finite baryon density have been performed
\cite{Boyd:1996bx,Karsch:2000ps,Fodor:2002sd}. These computations
are still limited to rather small lattices, therefore 
an alternative approach using an effective field theory is useful.
Previously, we have discussed a linear
$\sigma$-model with two massless quark flavours 
\cite {Schaefer:em,Meyer:1999bz} which is treated with
the renormalization group to capture the dynamics of the long-range
fluctuations near the critical point correctly. In
particular, this method includes dynamical chiral symmetry breaking as
the dominating feature of QCD at low energies.
The low-energy model lacks the gauge degrees of freedom, therefore it
cannot be linked with lattice calculations for $T \gtrsim 160$ MeV. 
As a first step, we propose to extend the flow equations of the linear $\sigma$-model 
by  a flow equation for
free massless quarks and gluons at high scale. Combining these two systems
within a single renormalization group flow  which integrates
successively over both high and low energy excitations one obtains 
an adequate equation of state of strong interaction over
a large range of  temperatures and densities, as a 
comparison of  our results with recent lattice simulations shows.

The $SU(2)_L \otimes SU(2)_R$ invariant
linear $\sigma$-model with constituent quarks $q$ and chiral mesons
$\Phi=(\sigma,\vec \pi)$ is an effective model for dynamical spontaneous chiral
symmetry breaking at intermediate scales of $0 \lesssim k
\lesssim \Lambda _M$. It is given 
by the Euclidean Lagrangian density
\begin{equation}
\label{lagrangian}
{\cal L}_{L\sigma M}=\bar q \slash{\partial} q +g \bar q
(\sigma+i\tau\pi\gamma_5) q+\tfrac{1}{2} (\partial_\mu
\Phi)^2+U(\Phi^2) \, . \nonumber
\end{equation}
The grand canonical partition function $Z$ for the quantum theory has
the path integral representation
\begin{align}
\label{partition-function}
Z&=\mathrm{Tr} \, e^{-\beta (H-\mu N)} \\ \nonumber
&=\int \! D \bar q \! \int \! D
q \! \int \! D \Phi \exp
\left( -\int_0^\beta \! d t_E \! \int \! d^3 x_E \, \left( {\cal L}_{L\sigma
      M}-\mu q^\dagger q \right) \right) , \nonumber
\end{align} 
where periodic and anti-periodic boundary conditions apply for bosons
and fermions, respectively. A Gaussian approximation to the path
integral followed by a Legendre transformation yields the 1-loop
effective action
\begin{align}   
  \Gamma[\Phi,\bar q, q]=&S^{\mathrm{uv}}[\Phi,\bar q, q] \!-\!
      \mathrm{Tr} \! \log \left( \frac{\delta^2 S[\Phi,\bar q,
      q]}{\delta \bar q(x) \delta q(y)} \right) \!+\!\tfrac{1}{2}
      \mathrm{Tr} \! \log \left( \frac{\delta^2 S[\Phi,\bar q,
      q]}{\delta \Phi^i(x) \delta \Phi^j(y)} \right) \, .
\end{align}
Here, the boundary conditions of the functional integral appear in
the momentum traces and we neglect contributions from
 mixed quark meson loops. We consider the effective
action $\Gamma$ in a local potential approximation
(LPA), which represents the lowest order in the derivative
expansion and incorporates fermionic as well as bosonic
contributions to the grand canonical potential density $\Omega$
which is related to the pressure of the system by
\begin{equation}
\label{PP}
\Omega(T,\mu)=-p(T,\mu) \, . \end{equation} \\
To derive RG flow equations, we use Schwinger's proper-time method to
regularize the respective logarithms.
A straightforward partial evaluation of the trace in momentum space
and a transformation of the expression to proper time
form\footnote{The transformation of the
  apparently complex expression to proper time form is possible, since
  all imaginary parts cancel each other in the Matsubara sum.} yields
\begin{align}
\Omega=&\Omega^{\mathrm{uv}}+\tfrac{1}{2} \int \frac{d \tau}{\tau}
      \, T \sum_{n=-\infty}^\infty \int \frac{d^{3} p}{(2\pi)^{3}}
      \left( 4 N_c N_f \text{e}^{-\tau \left( (\nu_n+i \mu)^2 + \vec
      p^2 + M_q^2 \right)} \right. \\
& \left. \qquad \qquad \qquad \qquad \qquad -3
      \text{e}^{-\tau \left(
      \omega_n^2 + \vec p^2 + M_\pi^2 \right)} - \text{e}^{-\tau
    \left( \omega_n^2 + \vec p^2 + M_\sigma^2 \right)} \right) f(\tau
      k^2) \nonumber \, .
\end{align}
where the Matsubara frequencies take the values $\omega_n=2 n \pi T$
for bosons and $\nu_n=(2 n+1) \pi T$ for fermions, respectively. The
effective masses are defined as 
\begin{equation}
M_q^2=g^2 \Phi^2 \; , \quad M_\pi^2=2 \, \frac{\delta \Omega} {\delta
  \Phi^2} \; , \quad M_\sigma^2=2 \, \frac{\delta \Omega }{\delta
  \Phi^2}+4 \, \Phi^2 \, \frac{\delta^2 \Omega}{(\delta \Phi^2)^2} 
\nonumber \, .
\end{equation}
A general set of smooth cutoff functions $f(\tau k^2)$, parametrized
by a real positive number $a$ fulfills the required
regularization conditions, cf. ref. \cite{Schaefer:em,Meyer:1999bz,Meyer:2001zp}
\begin{equation}
f^{(a)}(\tau k^2) \equiv \frac{\Gamma(a+1,\tau k^2)}{\Gamma(a+1)} \, .
\end{equation}
In previous works \cite{Schaefer:em,Meyer:1999bz,Meyer:2001zp}, cutoff
functions with integer $a$ have been applied to various
problems. However, in the case of a thermal system half-integer cutoff
functions are better suited, since they effectively regularize only the spatial
momentum integral. The relevant derivative
of the cutoff function 
with respect to the IR cutoff scale
\begin{equation}
k \frac{\partial f^{(a)}(\tau k^2)}{\partial
  k}=-\frac{2}{\Gamma(a+1)} (\tau k^2)^{a+1} \mathrm{e}^{-\tau k^2} \,
\nonumber
\end{equation}
involves a half-integer power of the proper time variable $\tau$,
which combines with the half integer power from the integration over
the spatial momentum coordinates. With $a=3/2$ the proper
time integral gives
an {\em analytic} RG improved
flow equation for the grand canonical potential $\Omega$ 
\begin{eqnarray}
\label{thermal-flow-equation}
k \frac{\partial \Omega _{L\sigma M} (\Phi^2,k)}{\partial k}&=&\frac{k^5}{6
    \pi^2} \left( \frac{3}{E_\pi} \left( \frac{1}{2}+n_B(E_\pi)
    \right) +\frac{1}{E_\sigma} \left( \frac{1}{2}+n_B(E_\sigma)
    \right) \right.  \nonumber \\ 
&& \qquad \qquad \qquad \quad \left. -\frac{2 N_c N_f}{E_q} \left(
    1-n_F(E_q)-\bar n_F(E_q) \right) \right) \, .
\end{eqnarray}
The appearing effective energies are defined by
\begin{equation}
E_i=\sqrt{k^2+M_i^2} \; , \quad i \in \{ \pi, \sigma, q \} \, , \nonumber
\end{equation}
and the occupation numbers have the usual form
\begin{equation}
n_B(E)=\frac{1}{e^{E/T}-1} \; , \quad n_F(E)=\frac{1}{e^{
    (E-\mu)/T}+1} \; , \quad \bar n_F(E)=\frac{1}{e^{(E+\mu)/T}+1}
    \, . \nonumber
\end{equation}
Note the exceptionally simple form of this evolution equation. One can
immediately read off the contributions from the vacuum and thermal
dynamics of the respective particles appearing with their proper
degeneracy factors. Especially, compared to previous vacuum
equations \cite{Meyer:2001zp} energy denominators appear instead of the
usual propagator terms, which is natural for a thermal equation
because of the broken Poincar\'e invariance. The analytic form of
these equations allows a numerical solution, especially in the 
case of finite temperature and density where the computational effort rises considerably. \\
The thermodynamics and phase structure of the chiral system has been
discussed in \cite{Schaefer:em,Meyer:1999bz,proceeding,Schaefer2}. One obtains a
good description of the second order chiral phase transition at $T_c
\approx 160$ MeV and also a qualitative picture of the first order phase
transition at high densities. However, the model lacks the
correct high-temperature behaviour since it does not contain the
appropriate degrees of freedom in this regime.


The fundamental degrees of freedom at high energies are quarks and
gluons of QCD, described by the Euclidean Lagrangian density
\begin{equation}
{\cal L}_{QCD}=\bar q \left( \slash{D} + m_c \right) q + \tfrac{1}{4}
F_{\mu \nu } F_{\mu \nu } \, , 
\end{equation}
where $m_c$ is the respective quark mass matrix. Due to asymptotic
freedom quarks and gluons decouple at
asymptotically high energies and the partition function
takes a similar form as Eq. (\ref{partition-function}) with chiral
mesons replaced by gluon fields.
Lattice gauge simulations, cf. ref. \cite{Boyd:1996bx,Karsch:2000ps}, 
suggest that  even at moderate
temperatures $T \gtrsim 200$ MeV the equation of state of the
interacting system behaves almost like a free gas.
Therefore in a first attempt to link the quark meson model with QCD 
we neglect the full gauge interactions and approximate the high-energy part of QCD at scales
$k>\Lambda_M$ by a free gas of
massless quarks and gluons. 
The integration of  the thermal
fluctuations within a common RG flow using the same smooth infrared cutoff
function allows a smooth transition of the
respective degrees of freedom. The matching depends on
the chosen cutoff function. A smooth cutoff function has the same
effect as a self-interaction of the gluons
and provides an effective screening for the gluonic degrees of freedom.
In order to match the two systems independently
of the matching scale one would have to add the relevant quark-gluon dynamics.
In this paper we show that, however, already the simple addition of the free quark gluon flow with the meson flow gives reasonable results for certain quantities as the equation of state. In further work we
plan to apply the evolution of the gluon system 
\cite{Gies:2002bs,Gies:2002hq,Meggiolaro:2000kp,Pirner:1991im,Arodz:mv} 
to reduce the dependence on the matching scale.

We neglect the current quark masses for the
two light quark flavors, then the equation for the high-energy part has
a very compact form in our simplified approximation:
\begin{eqnarray}
k \frac{\partial \Omega _{QCD} (k) }{\partial k} &=& 
\frac{k^4}{6
    \pi^2} \left( N_g ( \frac{1}{2}+n_B(k)
    ) -2 N_c N_f \left(
    1-n_F(k)-\bar n_F(k) \right) \right) \nonumber \\ &=&
k^4 \sum _{i \in \{
  g,q,\bar q\} } \frac{N_i}{6\pi
  ^{2}}\left(\frac{s_i}{2}+ n_i(k)\right) \, .
\end{eqnarray}
The sum runs over gluons, quarks and antiquarks 
including the appropriate occupation numbers with degeneracy factors and sign factors
$N_g=16$, $s_g\!=\!+1$ for the bosons and $N_q\!=\!N_{\bar q}\!=2
N_c N_f=\!12$,
$s_q\!=\!s_{\bar q}\!=\!-1$, for the fermions and antifermions, respectively. 
The potential of the total system is obtained by integrating  the
quark-gluon evolution equation from $k=\Lambda _{\infty}$
to $k=\Lambda _M$ and the quark-meson evolution equation from $k=\Lambda _M$
to $k=0$, i.e. the
IR-cutoff parameter of the quark-gluon potential is identical to the
UV-cutoff parameter of the quark-meson model. 
We can integrate the simplified QCD flow equation analytically
and get
\begin{align}
\label{QGflow}
&\Omega _{QCD} (\Lambda _{\infty},\Lambda _M,T)=\sum _{i} s_i
  \frac{N_{i}}{6\pi^{2}} \left[\
  \frac{k^{4}}{8}-k^{3}TLi_{1}\left(s_i n_{i}^{\infty}(k)\right)\right.\\ 
& \left.-3k^{2}T^{2}Li_{2}\left(s_i n_{i}^{\infty}(k)\right)-6k
  T^{3}Li_{3}\left(s_i n_{i}^{\infty}(k)\right)-6T^{4}Li_{4}\left(s_i
  n_{i}^{\infty}(k)\right) \right] _{\Lambda _{\infty}} ^{\Lambda _M} \nonumber \, . 
\end{align}

The polylogarithmic functions $\text{Li}_n (x)$ are
defined by
 \begin{equation}
Li_{1}(x)=-\ln (1-x)\; ,\quad Li_{n}(x)=\int
_{0}^{x}\frac{Li_{n-1}(z)}{z}dz \nonumber
\end{equation}
and the functions $n_{i}^{\infty}(k)$ denote the classical
Maxwell-Boltzmann distribution functions of the respective particles
\begin{equation}
n_{g}^{\infty}(k)=e^{-\frac{k}{T}} \; , \quad
n_{q}^{\infty}(k)=e^{-\frac{k -\mu}{T}} \; , \quad
n_{\bar{q}}^{\infty}(k)=e^{-\frac{k +\mu}{T}} \nonumber\, .
\end{equation}
We can evaluate eq. (\ref{QGflow}) in the limit $\Lambda
_{\infty}\rightarrow \infty$ by making use of the behaviour of the
polylogarithms in the vicinity of zero
\begin{align}
\label{QGpot}
&\Omega _{QCD} (\Lambda _{\infty},\Lambda _M,T)=\sum _{i} s_i
  \frac{N_{i}}{6\pi^{2}} \left(
  -\frac{\Lambda_{\infty}^{4}}{8}+\frac{\Lambda_M ^{4}}{8}-\Lambda _M
  ^{3}TLi_{1}(s_i n_{i}^{\infty}(\Lambda _M))\right.\\ 
& \left.-3\Lambda _M ^{2}T^{2}Li_{2}(s_i n_{i}^{\infty}(\Lambda
  _M))-6\Lambda _M
  T^{3}Li_{3}(s_i n_{i}^{\infty}(\Lambda _M))-6T^{4}Li_{4}(s_i
  n_{i}^{\infty}(\Lambda _M)) \right) \nonumber \, .
\end{align}
Here we kept the divergent part proportional to $\Lambda _{\infty}
^{4}$, so the potential still needs to be regularized.
In the limit of zero temperature the vacuum part can be
identified, since it is only a function 
of $\Lambda _M$ and $\Lambda _{\infty}$
\begin{equation}
\label{QGvac}
\Omega ^{vac} _{QCD} (\Lambda_{\infty },\Lambda_M,T=0 )=\sum _{i}
  s_i\frac{N_{i}}{6\pi^{2}} \left( \frac{\Lambda_M ^{4}}{8}
  -\frac{\Lambda_{\infty}^{4}}{8} \right) \, . 
\end{equation}

\begin{figure}[hbt]
\begin{center}
\begin{raggedright}
\epsfig{file=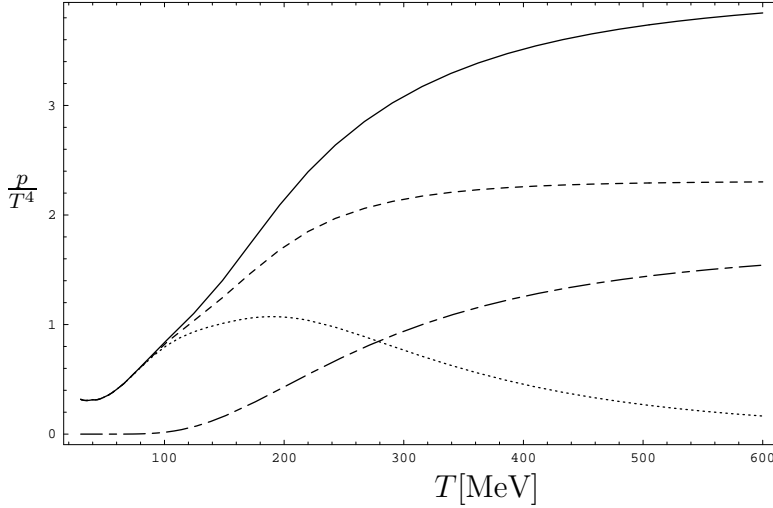,width=10cm}
\end{raggedright} 
\flushleft \vspace{-0.1cm} \hspace{7.1cm} {$T [\mbox{MeV}]$}
\flushleft \vspace{-4.5cm} \hspace{1.4cm} {$\frac{p}{T^4}$}
\vspace{3.7cm}
\caption{\label{fig:presscontr} Different contributions to
  $\frac{p}{T^4}$ arising from the mesons and quarks (constituent
  quarks and free quarks )
  (dashed line) and the gluons (dash-dotted line). The
  dotted line shows the scaled pressure function of the linear
  $\sigma$-model with two quark flavours which contains the cutoff $\Lambda_M$.}
\end{center}
\end{figure}
To obtain the potential
$\Omega _{L\sigma M}$ of the chiral low-energy part, we integrate the 
corresponding flow equation from $k=\Lambda _M$ to k=0. As initial values
we choose $\Lambda_M =1.0$ GeV,
$U(\Phi^2)=\frac{1}{2}m ^2 _{M}\Phi ^2 +\frac{1}{4}\lambda _{M}\Phi
^4$  with $\lambda _{M} =37$, $m _{M}=320$ MeV and the quark-meson
coupling $g=3.45$. These initial parameters are adjusted to get the value of
the pion decay constant $<\sigma>=f_\pi =88$ MeV at k=0 in the chiral limit.
In principle, the starting parameters should result from the
integration over the QCD degrees of freedom. This may be possible 
including quark interactions from gluons at higher scales.

The full potential reads:
\begin{equation}
\Omega_{total} (\Lambda_{\infty},k=0,T)= \Omega_{QCD}(\Lambda
  _{\infty},\Lambda _M,T)+\Omega_{L \sigma M}(\Lambda_M,k=0,T)\, .
\end{equation} 
The finite-temperature part relevant for the pressure does not
depend on the vacuum energy. Therefore in the regularized potential the
UV-divergent part drops out
as it must
\begin{equation}
p(T)= -\Omega_{total} (\Lambda_{\infty},k=0,T)+\Omega_{total}
(\Lambda_{\infty},k=0,T=0)\, .
\end{equation}
In the limit of infinite temperature  $T \rightarrow \infty$ the ratio of $p/T^4$ yields the
Stefan-Boltzmann limit 
for a massless quark-gluon gas, because the region $k<\Lambda_M$ 
has vanishing weight
\begin{equation}
\frac{p(T)}{T^4} \rightarrow \left( N_{g}+\frac{7}{8} \left( N_{q}+N_{\bar{q}}
  \right) \right)\frac{\pi^{2}}{90}\, .
\end{equation}

For small temperatures $T \rightarrow 0$ we find a massless gas of pions with 
\begin{equation}
\frac{p(T)}{T^4} \rightarrow 3\frac{\pi^{2}}{90}\, .
\end{equation}

In fig. \ref{fig:presscontr} we show the different contributions
to $p/T^4$ arising from the integration over the flow of the 
mesons and all quark contributions (including both the constituent
quarks with cutoff $\Lambda_M$ and free quarks without cutoff)(dashed
line) 
and the integration over the 
gluons (dash-dotted line). Furthermore we plot the scaled
pressure of the linear $\sigma$-model with cutoff $\Lambda_M$ 
(dotted line) described by
eq. (\ref{thermal-flow-equation}). 

The comparison indicates clearly
that the quark-meson model with cutoff $\Lambda_M$ alone becomes unreliable
for a description of the equation of state at temperatures $T>150$ MeV.
It definitely has to be linked with quark and gluon degrees of freedom
at higher temperatures.
The linear sigma model is adequate at low temperature where 
massless pions dominate the pressure and the quarks are massive 
due to spontaneous breaking of the chiral symmetry.
Gluons contribute to the pressure only at rather high
temperatures since the smooth infrared cutoff around the scale $\Lambda_M$ introduces
a gluon mass and gives deviations between our model and the free gluon gas. The
flow equation with the successive integration over different degrees
of freedom avoids a first-order transition which typically arises in
simplified models with two components like the bag model combined
with a pion gas \cite{Chodos:1974je}.

\begin{figure}[t]
\begin{center}
\begin{raggedright}
\epsfig{file=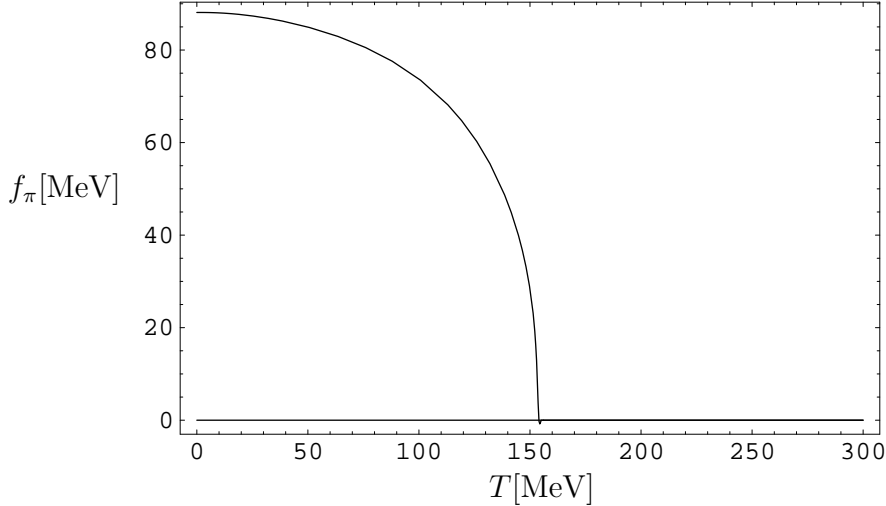,width=10cm}
\end{raggedright} 
\flushleft \vspace{-0.1cm} \hspace{6.5cm} {$T [\mbox{MeV}]$}
\flushleft \vspace{-4.5cm} \hspace{0.1cm} {$f _{\pi} [\mbox{MeV}]$}
\vspace{3.7cm}
\caption{\label{fig:fpi}.The pion decay constant as a function of temperature
  for vanishing chemical potential.}
\end{center}
\end{figure}

The critical behaviour of the system is described by the pion decay constant
which is shown as a function of temperature for vanishing chemical potential
in fig. \ref{fig:fpi}. The figure shows that the pion decay constant goes to
zero for $T _c=154$~MeV continuously. For temperatures below $T _c$ the quarks
are massive and the pions are massless, so the system is in the chiral broken
phase. For temperatures above $T _c$ the system is in the chiral symmetric
phase, where the quarks are massless and the pions
are massive. So one obtains a second order chiral phase
transition at $T= T_c$, which is about $20$ MeV smaller than the 
result of lattice calculations \cite{Karsch:2000ps}. Note that the pion decay
constant as a function of temperature for the chiral low energy system does
not differ from the function of the total system, because the pion decay constant is zero in the
high-energy part of our model.

\begin{figure}[t]
\begin{center}
\begin{raggedright}
\epsfig{file=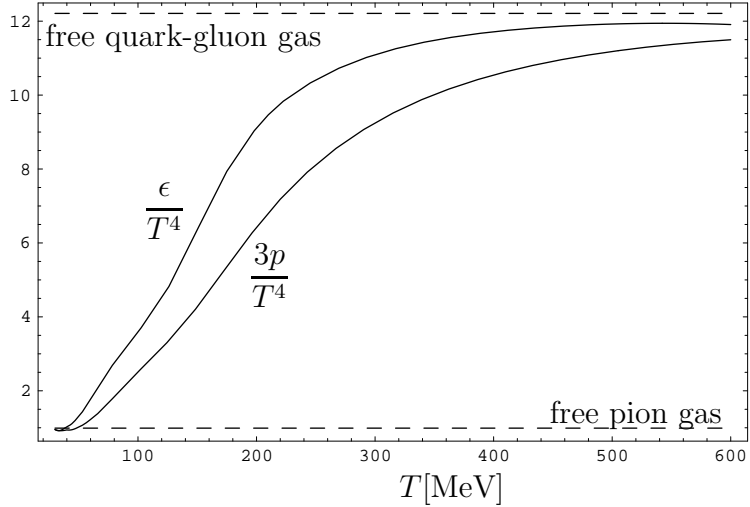,width=10cm}
\end{raggedright} 
\flushleft \vspace{-0.1cm} \hspace{7.1cm} {$T [\mbox{MeV}]$}
\flushleft \vspace{-1.5cm} \hspace{9.1cm} {free pion gas}
\flushleft \vspace{-2.5cm} \hspace{5.1cm} {${\displaystyle \frac{3p}{T^4}}$}
\flushleft \vspace{-1.7cm} \hspace{3.7cm} {${\displaystyle \frac{\epsilon}{T^4}}$}
\flushleft \vspace{-2.9cm} \hspace{2.4cm} {free quark-gluon gas}
\vspace{5.9cm}
\caption{\label{fig:scaledpressure} The scaled pressure density multiplied by
  a factor of three compared to the scaled energy density as a
function of temperature. The upper and lower Stefan-Boltzmann 
limits corresponding to a free quark-gluon gas and a pion
  gas are drawn as dashed lines.}
\end{center}
\end{figure}

In fig. \ref{fig:scaledpressure} we compare the curve of the scaled
energy $\frac{\varepsilon}{T^4}$ with three times 
the pressure $\frac{ 3 p}{T^4}$ as a
function of temperature. Both quantities
$\frac{\varepsilon}{T^4},\frac{ 3 p}{T^4}$
gradually increase across the second order phase transition. The
scaled pressure density reaches about $95$\% of the Stefan-Boltzmann
limit at $T \approx 4\ T_c$. Respective lattice calculations
\cite{Karsch:2000ps} reach about $85$\% of the Stefan-Boltzmann limit
at $T \approx 4\ T_c$. 
The energy density becomes more rapidly asymptotic, cf. fig. \ref{fig:scaledpressure}. 
Thereby, the ``interaction measure'' $\varepsilon -3
p$ has a peak at $T \approx 1.27\, T_c \approx 196$ MeV, whereas the
respective maximum of lattice calculations of (2+1)-flavour QCD
\cite{Fodor:2002sd} is located at approximately $T\approx 1.1\, T_c
\approx 189 $ MeV. The ``interaction measure'' $\varepsilon -3 p$ has more subtle features at finite chemical potential. For instance we find
a low temperature maximum in  $\varepsilon -3 p$
at finite chemical potential which does not agree with 
finite density lattice calculations.
This low temperature maximum arises 
from the fact that we have constituent quarks in
our model and no baryons for temperatures below $T_c$. 
Lattice calculations of the same quantity go rapidly to
zero for temperatures below $T_c$ which may be an effect of
confinement and/or of the limited grid sizes and the high masses of the nucleons.

\begin{figure}[t]
\begin{center}
\begin{raggedright}
\epsfig{file=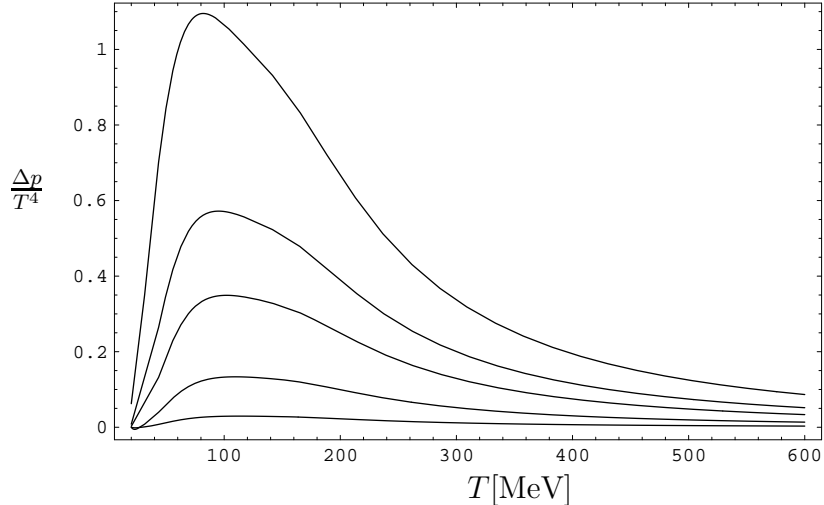,width=10cm}
\end{raggedright} 
\flushleft \vspace{-0.1cm} \hspace{7.1cm} {$T [\mbox{MeV}]$}
\flushleft \vspace{-4.5cm} \hspace{1.0cm} {$\frac{\Delta p}{T^4}$}
\vspace{3.7cm}
\caption{\label{fig:deltap} The scaled pressure difference 
$\frac{\Delta p}{T^4} = 
\frac {p(T,\mu)-p(T,\mu=0)}{T^4}$ is shown as a funtion of temperature for $\mu
  _B=100,210,330,410,530$ MeV (from bottom to top).}
\end{center}
\end{figure}

To study the effects of a finite baryon chemical potential on
the pressure it is useful to consider the difference $\Delta p$ of the
pressure of the baryonic system to the nonbaryonic system at the same
temperature:
\begin{equation}
\Delta p= (p(T,\mu)-p(T,\mu=0))\, .
\end{equation} 
Small chemical potentials compared to the baryon mass suppress the
baryonic pressure at low temperatures. Our calculation shows a
suppression of $\Delta p /T^4$ in agreement with the low-temperature expansion for a gas of
nonrelativistic and non-interacting constituent quarks with masses 
$m>\mu=\mu_B/3$, which reads 
\begin{equation}
\frac {\Delta p}{T^4}=\frac{g}{2\sqrt{2}\pi
  ^{\frac{3}{2}}}\left(\frac{m}{T}\right)^{\frac{3}{2}}\left(e^{-\frac{m-\mu}{T}}-e^{-\frac{m}{T}}\right)
  \, .\label{Pentw}
\end{equation}
Here $g$ denotes the degeneracy factor of the fermions. In fig. \ref{fig:deltap} one can see the scaled pressure $\frac{\Delta p}{T^4}$
of the quarks for different
chemical potentials $\mu_B=100$ MeV to $\mu_B=530$ MeV. We have chosen 
these values identical to the ones in the lattice calculation in
ref. \cite{Fodor:2002sd}, where the equation of state includes heavy
strange quarks.
At high temperatures the scaled pressure difference decreases according to
\begin{equation}
\frac {\Delta p}{T^4} =4N_c N_f \left(\frac{\mu^2}{24 T^2} +\frac{\mu
^4}{48\pi ^2 T^4}\right) \, . \label{Phoch}
\end{equation}
This behaviour is also reproduced by our calculation, cf. fig. \ref{fig:deltap}.

Linking the quark-meson model with QCD degrees of freedom we find
that the low-energy sigma model can be extended to high temperature
$T>150$ MeV. The combination of different degrees of freedom within
the renormalization group flow equations opens the window for a 
consistent description of QCD both at low and high scales which
is needed in the equation of state. Already the simplified model we
constructed shows good agreement with lattice simulations.
As indicated above, the low-momentum behaviour of the model
has to be improved to get  better agreement with the physics for
temperatures below $T _c$.  Furthermore the matching
between the quark-gluon and quark-meson model is too simple
and should be extended by including the relevant quark-gluon dynamics on the high-energy
side. 
Both the  effective four-fermion coupling and gluon
condensation depend on the matching scale. 
A scale independent transition may be achieved
if one can connect the chiral symmetry dynamics with 
gluon condensation.

\end{document}